\documentclass[11pt,twoside,dvips]{article}
\usepackage{verbatim}
\usepackage{amssymb}
\usepackage{amsfonts}
\usepackage{amsmath}
\usepackage{pifont}
\usepackage[spanish,english]{babel}
\usepackage[pdftex]{graphicx}
\usepackage{verbatim}
\usepackage{epsfig}
\usepackage{bm}
\usepackage{longtable}
\usepackage{fancyhdr}
\begin{document}

\fancypagestyle{plain}{%
\fancyhf{}%
\fancyhead[LO, RE]{XXXVIII International Symposium on Physics in Collision, \\ Bogot\'a, Colombia, 11-15 September 2018}}

\fancyhead{}%
\fancyhead[LO, RE]{XXXVIII International Symposium on Physics in Collision, \\ Bogot\'a, Colombia, 11-15 September 2018}

\title{The Future of High-Energy Collider Physics}
\author{John Ellis$\thanks{%
e-mail: John.Ellis@cern.ch}$
\\ Physics Department, Kings College London, Strand, London WC2R 2LS, UK; \\
Theoretical Physics Department, CERN, CH-1211 Geneva 23, Switzerland; \\
National Institute of Chemical Physics \& Biophysics, R\"avala 10, 10143 Tallinn, Estonia}
\date{}
\maketitle

\begin{abstract}
High-energy collider physics in the next decade will be dominated by the LHC, whose high-luminosity incarnation will take Higgs measurements
and new particle searches to the next level. Several high-energy $e^+ e^-$ colliders are being proposed, 
including the ILC (the most mature), CLIC (the highest energy)
and the large circular colliders FCC-ee and CEPC (the highest luminosities for $ZH$ production, $Z$ pole and $W^+ W^-$ threshold studies), and
the latter have synergies with the 100-TeV $pp$ collider options for the same tunnels (FCC-hh and SppC). The Higgs, the Standard Model effective field
theory, dark matter and supersymmetry will be used to illustrate some of these colliders' capabilities. Large circular colliders appear the most versatile, able to
explore the 10-TeV scale both directly in $pp$ collisions and indirectly via precision measurements in $e^+ e^-$ collisions.
\end{abstract}

\begin{center}
KCL-PH-TH/2018-61, CERN-TH/2018-228
\end{center}

\section{Introduction}

At the time of writing, the Standard Model (SM) still reigns rather supreme. Calculations in its QCD sector are becoming ever
more precise, and LHC data are highly compatible with its predictions. Electroweak precision tests are also quite consistent
with the SM, {\it modulo} longstanding issues such as $Z$ decays to $b$ quarks and the anomalous magnetic moment of the
muon. There are some anomalies in $B$ meson decays, but more data are required before any conclusions can be reached.
The Higgs boson discovered in 2012 continues obstinately to obey the predictions of the SM.

How will we discover physics beyond the SM? Electroweak, Higgs and flavour data may provide clues or evidence.
So might astrophysics and cosmology, e.g., via measurements of cosmic rays or gravitational waves. However, just as
previous collider experiments pinned down the SM, future collider experiments with surely be needed to pin down the
physics beyond the SM. In this talk I present a personal review of planned and projected future colliders and illustrate how they might take us
beyond the SM, using Higgs studies, the Standard Model effective field theory, dark matter and supersymmetry as illustrations.

\section{Possible Future Colliders}

\subsection{The High-Luminosity LHC}

Following Run 2 of the LHC there is a 2-year shutdown for refurbishment and upgrades, following which Run~3 should
accumulate about 300/fb of integrated luminosity in each of ATLAS and CMS. After another, longer shutdown for
the ultimate upgrades of the accelerator and experiments, the high-luminosity LHC (HL-LHC) will operate for about a
decade from 2026 onwards, with an integrated luminosity target of 3000/fb, almost two orders of magnitude more than
the luminosities used in most current LHC analyses~\cite{HL-LHC}.

\begin{figure}[h!]
\centering
\includegraphics[scale=0.4]{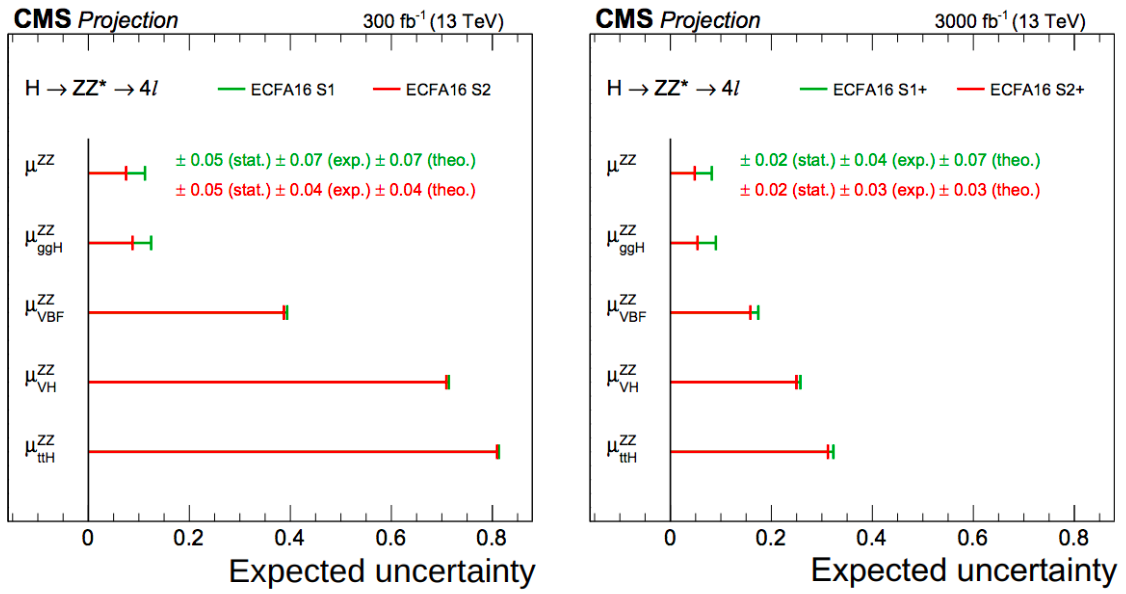}
\caption{\it The prospective accuracies of CMS measurements of $H \to ZZ^*$ with 300 (3000)/fb of
data in the left (right) panel, under less (more) optimistic assumptions about the attainable systematic
and theoretical uncertainties in green (red)~\cite{Scott}. }
\label{fig:HL-LHCH}
\end{figure}

The increase in luminosity will make possible much more precise measurements of the properties of the Higgs boson,
as illustrated in Fig.~\ref{fig:HL-LHCH}~\cite{Scott}. As seen there, for the full benefits of the increased luminosity to be realized,
it will be necessary to reduce the systematic uncertainties, hopefully as $1/\sqrt{L}$, and also to halve
the theoretical uncertainties, requiring higher-order QCD and electroweak corrections in many cases. HL-LHC will
also increase significantly the reach for new massive particles such as those predicted by supersymmetry.

The HL-LHC project has been fully approved by the CERN Council and will operate until about 2 decades hence.
However, in light of the long lead times for proposing, preparing and constructing new high-energy colliders, there is
much ongoing discussion about possible future colliders.

\subsection{Possible Future $e^+ e^-$ Colliders}

\begin{figure}[h!]
\centering
\includegraphics[scale=0.3]{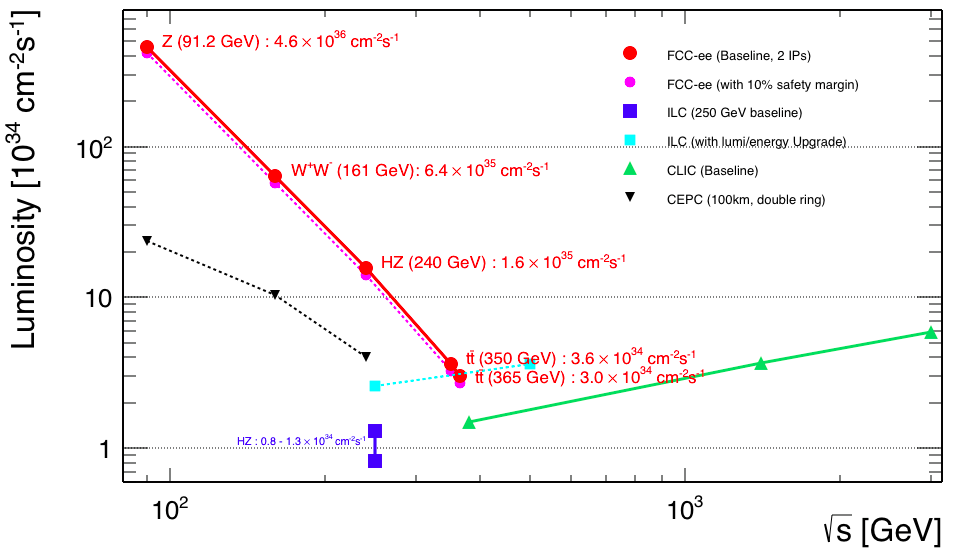}
\caption{\it The prospective centre-of-mass energy reaches and luminosity targets of high-energy $e^+ e^-$
collider projects.}
\label{fig:e+e-ColliderL}
\end{figure}

In particular, there is much discussion of different possible future high-energy $e^+ e^-$ colliders. whose
potential energy ranges and luminosities are shown in Fig.~\ref{fig:e+e-ColliderL}. The most mature of these
projects is the ILC, which is currently proposed to run at 250~GeV in the centre of mass, with possible upgrades
to 500~GeV or 1~TeV. The highest-energy option is CLIC, which aims to reach 3~TeV in the centre of mass.
The circular $e^+ e^-$ colliders FCC-ee and CEPC are limited to centre-of-mass energies $\lesssim 400$~GeV,
but could attain higher luminosities at the $ZH$ and $W^+ W^-$ thresholds, as well as at the $Z$ peak.

\begin{figure}[h!]
\centering
\includegraphics[scale=0.3]{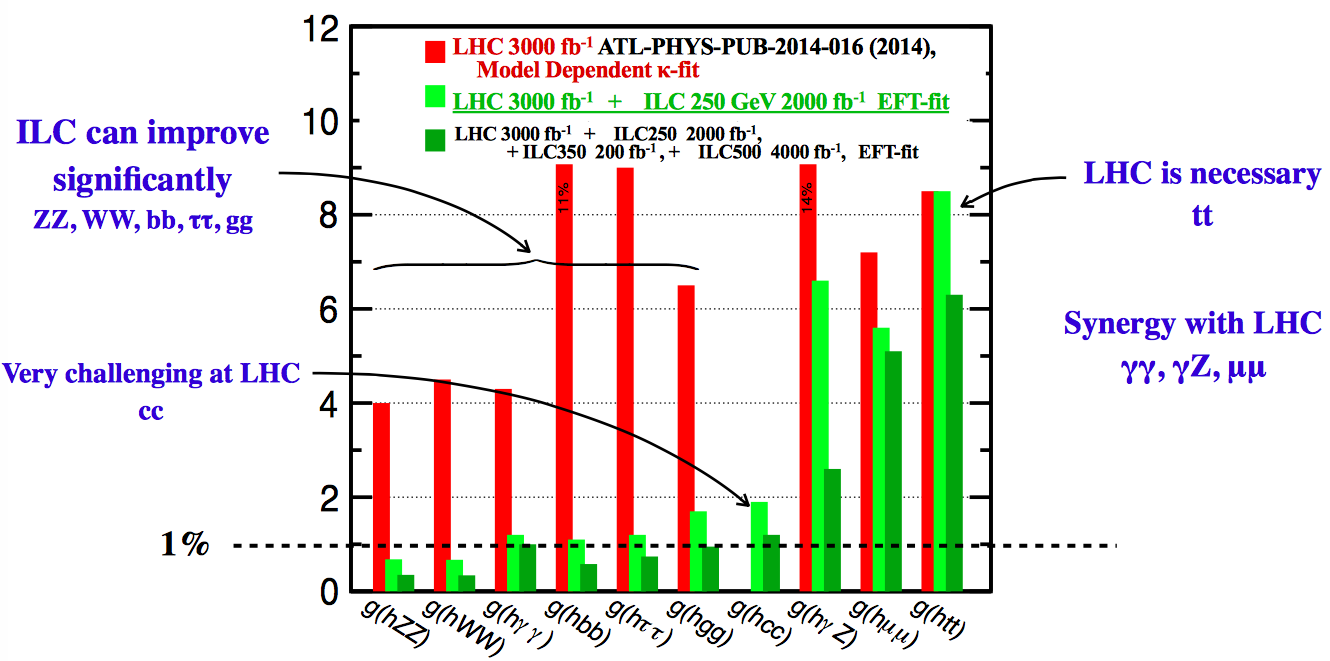}
\caption{\it Complementarities between prospective Higgs coupling measurements at HL-LHC and the ILC~\cite{Ogawa}.}
\label{fig:complement}
\end{figure}

The ILC is currently awaiting approval by the Japanese government, and  its prospective measurements of
Higgs couplings would complement those possible with HL-LHC, as seen in Fig.~\ref{fig:complement}~\cite{Ogawa}. On
the other hand, the absence of new particles so far at the LHC has diminished the chance that the ILC could
discover any, and the higher luminosity of a circular $e^+ e^-$ collider would enable it to make more accurate
measurements of the Higgs boson, as discussed below.

\subsection{Possible Circular Colliders}

The vision being developed in Europe (FCC)~\cite{FCC,FCCee} and China (CepC/SppC)~\cite{CEPC} is of a $\sim$ circular tunnel of 
$\sim 100$~km circumference that could house an $e^+ e^-$ collider, a $pp$ collider using magnets 
with fields of 16 to 20 Tesla that would be capable of reaching $\sim 100$~TeV in the centre of mass,
and an $e p$ collider. Such a complex would be capable of exploring the 10-TeV scale both directly in $pp$
collisions and indirectly in $e^+ e^-$ and $e p$ collisions. Another possibility is to replace the present LHC 
magnets with similar high-field magnets so as to reach $\sim 27$~TeV, the HE-LHC~\cite{HE-LHC}.

\begin{figure}[h!]
\centering
\includegraphics[scale=0.35]{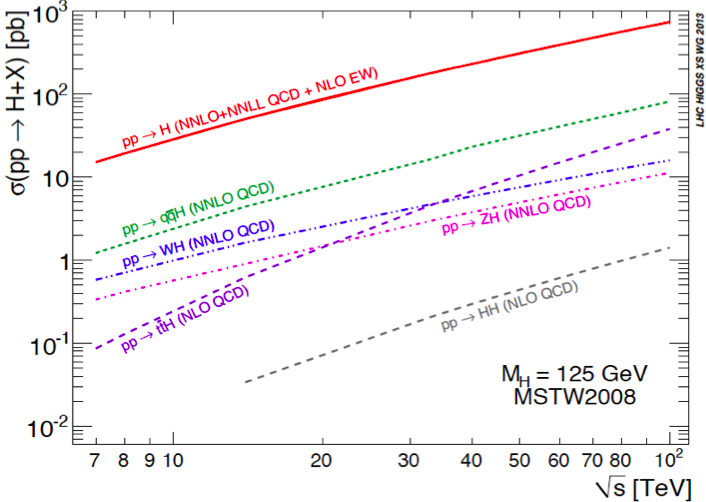}
\caption{\it The cross-sections for various Higgs production mechanisms at $pp$ centre-of-mass energies
$\lesssim 100$~TeV~\cite{LHCHXSWG}.}
\label{fig:HXS}
\end{figure}

The cross-sections for various Higgs production mechanisms at high-energy $pp$ colliders are shown in
Fig.~\ref{fig:HXS}. We see that many of the cross-sections increase by almost two orders of magnitude
between the LHC and a 100-TeV collider~\cite{LHCHXSWG}. This is the case, in particular, for double Higgs production. We note
also that HE-LHC and FCC-pp have target luminosities up to $3 \times 10^{35}$/cm$^2$s, a factor $\sim 5$
larger than HL-LHC.

\begin{figure}[h!]
\centering
\includegraphics[scale=0.25]{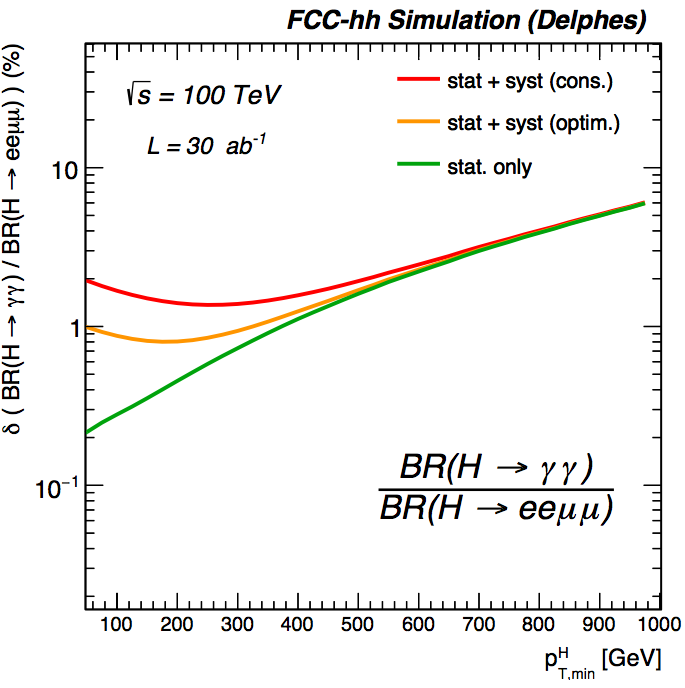}
\includegraphics[scale=0.45]{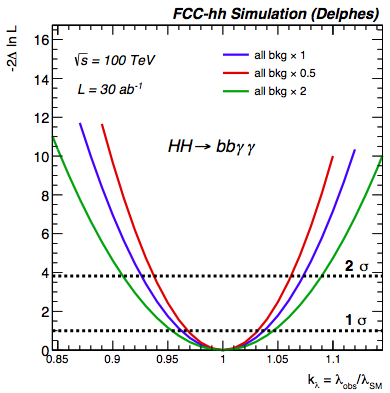}
\caption{\it The potential accuracies of FCC-hh measurements of the ratio of $H \to \gamma \gamma$ and 
$Z Z^*$ branching ratios (left panel)~\cite{Selvaggi} and of the triple-Higgs coupling (right panel)~\cite{OrtonaSelvaggi}.}
\label{fig:FCChhH}
\end{figure}

These high cross-sections and luminosities will make possible very accurate measurements of Higgs production,
of which a couple of examples are shown in Fig.~\ref{fig:FCChhH}. We see in the left panel that a $\sim 1$\%
measurement of the ratio of $H \to \gamma \gamma$ and $Z Z^*$ branching ratios should be possible at low
$p_T$ (with an optimistic estimate of the systematic uncertainties), and even a $\sim 10$\% measurement at 
$p_T \sim 1$~TeV~\cite{Selvaggi,FCCppH}. The right panel of Fig.~\ref{fig:FCChhH} shows that a $\sim 5$\% measurement of the
triple-Higgs coupling should be possible with FCC-hh~\cite{OrtonaSelvaggi,FCCppH}.

\begin{figure}[h!]
\centering
\includegraphics[scale=0.22]{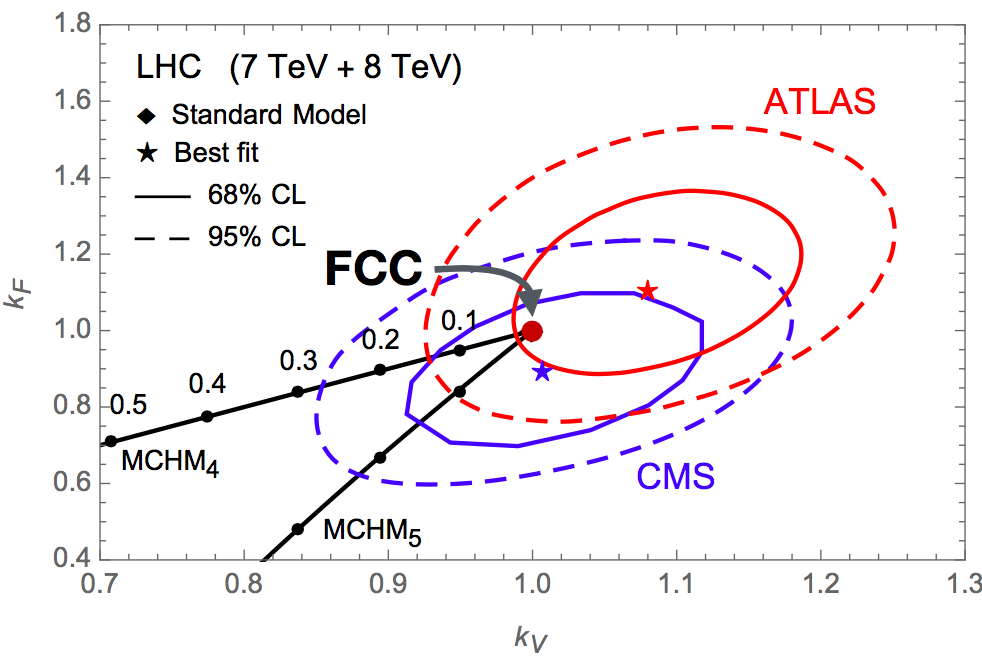}
\includegraphics[scale=0.36]{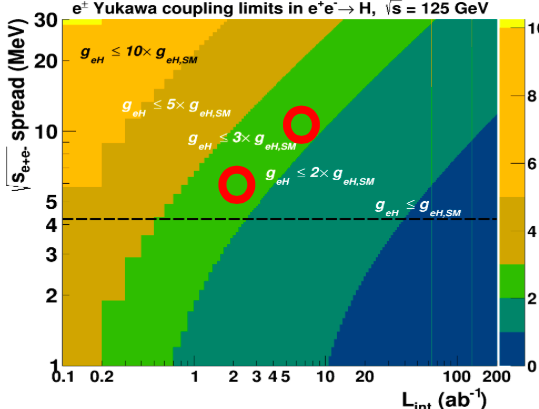}
\caption{\it The potential accuracies of combined FCC-ee and FCC-hh measurements of the Higgs coupling
modifiers $\kappa_{V,F}$ (left panel)~\cite{Wulzer}, and FCC-ee could provide interesting constraints on the H coupling to electrons (right panel)~\cite{Hee},
under suitable running conditions (red circles).}
\label{fig:FCCH}
\end{figure}

The combination of results from $e^+e^-$ and $pp$ collisions (FCC-ee/FCC-pp or CEPC/SppC) would provide
important synergies and the greatest precision in measurements of the Higgs couplings. The left panel of 
Fig.~\ref{fig:FCCH} shows the precision in the Higgs coupling modifiers $\kappa_{V,F}$ attainable by combining
FCC-ee and FCC-pp measurements, compared with the current accuracies of LHC measurements~\cite{Wulzer}. Also
shown are predictions of two composite Higgs models, MCHM$_{4,5}$: the present measurements constrain
their parameters at the 10 to 20\% level, whereas the future measurements could take the accuracy down to
a fraction of 1\%. The LHC has already measured the Higgs couplings to the third-generation fermions $t, b, \tau$,
the coupling to muons is within its reach, and $H \to c \bar c$ could be measured at an $e^+ e^-$ collider. 
However, measuring the Higgs couplings to $u, d, s$ quarks and to electrons remain
dreams, though the right panel of Fig.~\ref{fig:FCCH} shows that with a suitable running scheme
FCC-ee measurements might be able to reach with a factor $\sim 2.2$ of the SM value of the $H e^+ e^-$ coupling at the 95\% CL~\cite{Hee}.

\section{Standard Model Effective Field Theory}

In the previous Section various future collider projects were introduced, and their capabilities illustrated
via possible Higgs measurements. In this Section I discuss a broader approach to collider physics, based
on the Standard Model Effective Field Theory (SMEFT). In this approach one looks systematically for possible
new high-mass physics beyond the SM via the higher-dimensional effective interactions that it might induce
between SM particles. The most relevant for collider physics are generated by operators of dimension 6
whose coefficients are scaled by $1/\Lambda^2$, where $\Lambda$ is the scale of new physics, so the first
step is to classify these operators. Assuming that their coefficients are flavour-independent, just 20 such
operators are important for precision electroweak data, diboson and Higgs production. Since their
contributions to amplitudes are $\propto 1/\Lambda^2$, their relative importance increases with the
centre-of-mass energy and/or transverse momentum. This gives an advantage to a higher-energy $e^+ e^-$
collider such as CLIC, as we see later, and renders kinematic measurements particularly useful. 

\begin{figure}[h!]
\centering
\includegraphics[scale=0.4]{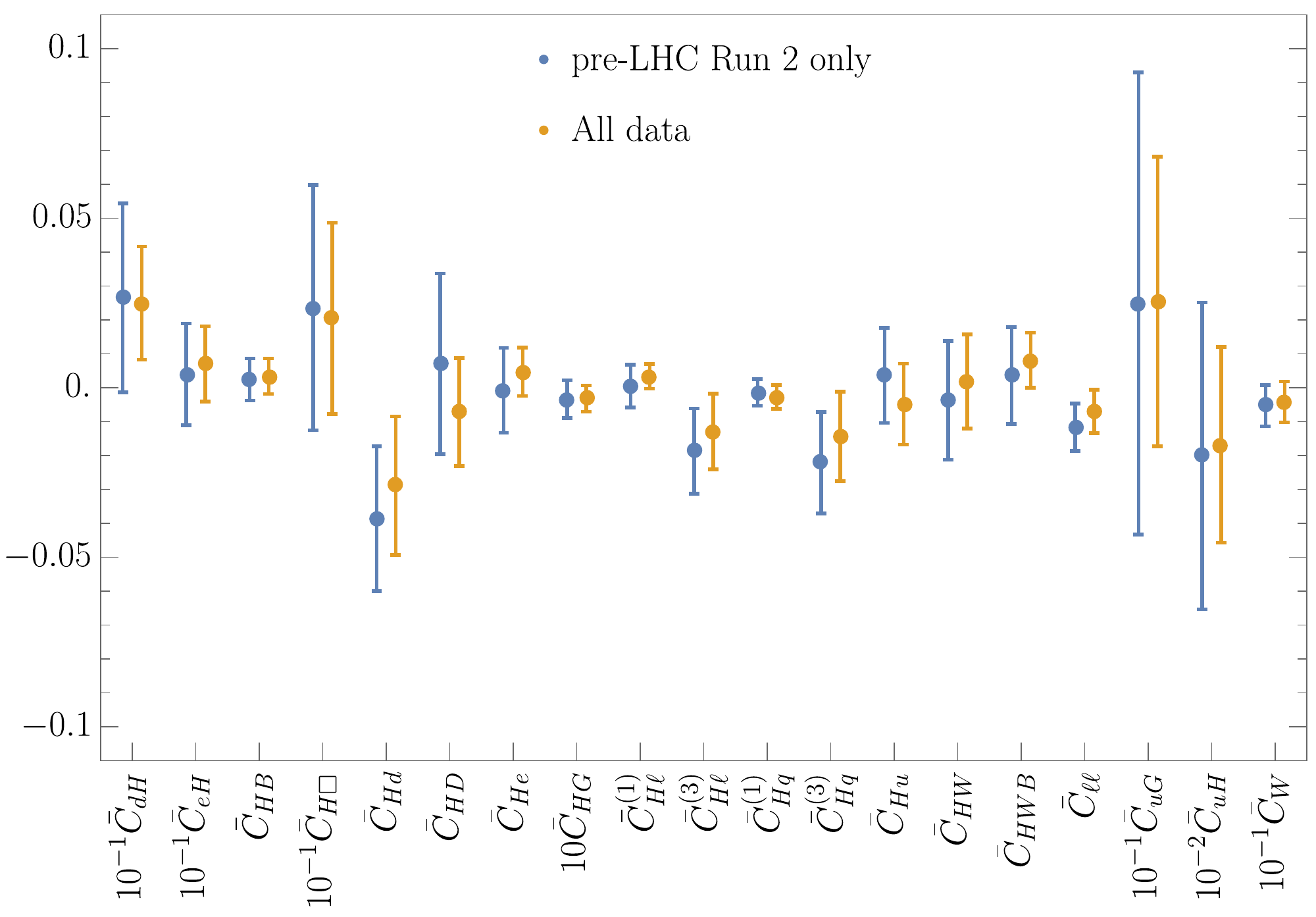}
\includegraphics[scale=0.4]{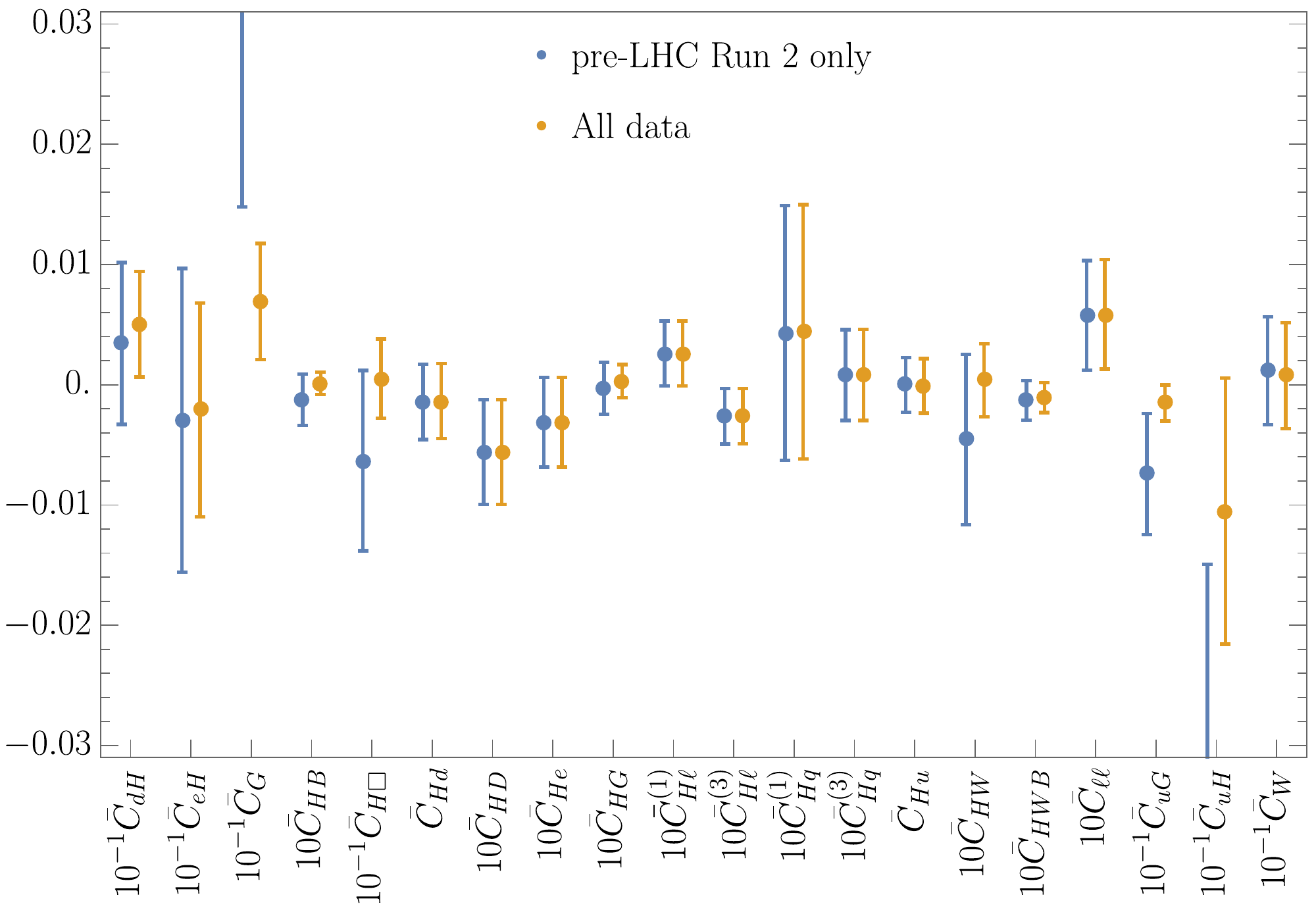}
\caption{\it Best-fit values and 95\% CL ranges of operator coefficients from global fits (orange) including all operators 
simultaneously (left panel) and switching each operator on individually (right panel)~\cite{EMSY}. 
Also shown are fits omitting the LHC Run 2 data (blue).}
\label{fig:SMEFT}
\end{figure}

We recently made a global SMEFT fit to all the available Higgs data from LHC Runs 1 and 2 including a 
number of kinematic measurements from ATLAS, the most
sensitive high-$p_T$ $W^+ W^-$ measurement from the LHC, and the precision electroweak and diboson 
production data from LEP~\cite{EMSY}. Constraints on operator coefficients from this fit are shown in Fig.~\ref{fig:SMEFT}.
The blue ranges were obtained using only pre-Run-2 data, whereas the orange ranges include all the data,
The left panel shows marginalized results when all operators are allowed to contribute, and the right panel 
shows results when each operator is switched on individually. As could be expected, the fits to individual give,
in general, tighter constraints than all operators are included and marginalized. We also see that
the fit using Run 2 data gives significantly tighter constraints on many operator coefficients.

\begin{figure}[t!]
  \centering
\includegraphics[width=0.5\textwidth]{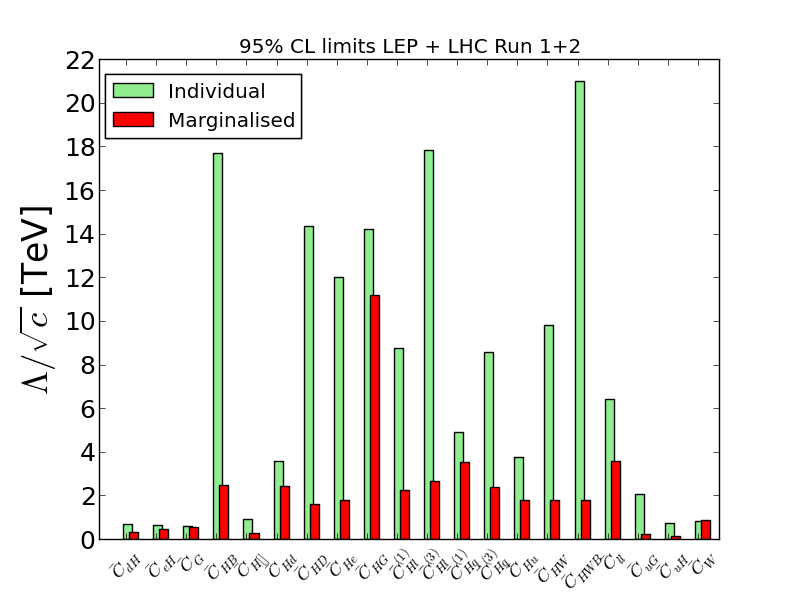}
 \caption{\it Comparison of the 95\% CL bounds obtained from marginalized (red) and individual (green) fits to the 
 20 dimension-6 operators contributing to electroweak precision tests, diboson and Higgs measurements~\cite{EMSY}.}
   \label{fig:bars}
\end{figure}

A direct comparison between the 95\% CL constraints on operator coefficients in marginalized and individual fits
to all the available data is shown in Fig.~\ref{fig:bars}~\cite{EMSY}. As before, we see that the individual fits give significantly
tighter constraints on most operator coefficients, extending in many cases to values of $\Lambda$ in the multi-TeV 
range for coefficients $c = {\cal O}(1)$. This analysis reveals no hint of physics beyond the SM, since the SMEFT
fit has a global $p$-value that is no better than the SM.

\begin{figure}
  \centering
  \includegraphics[width=0.45\textwidth]{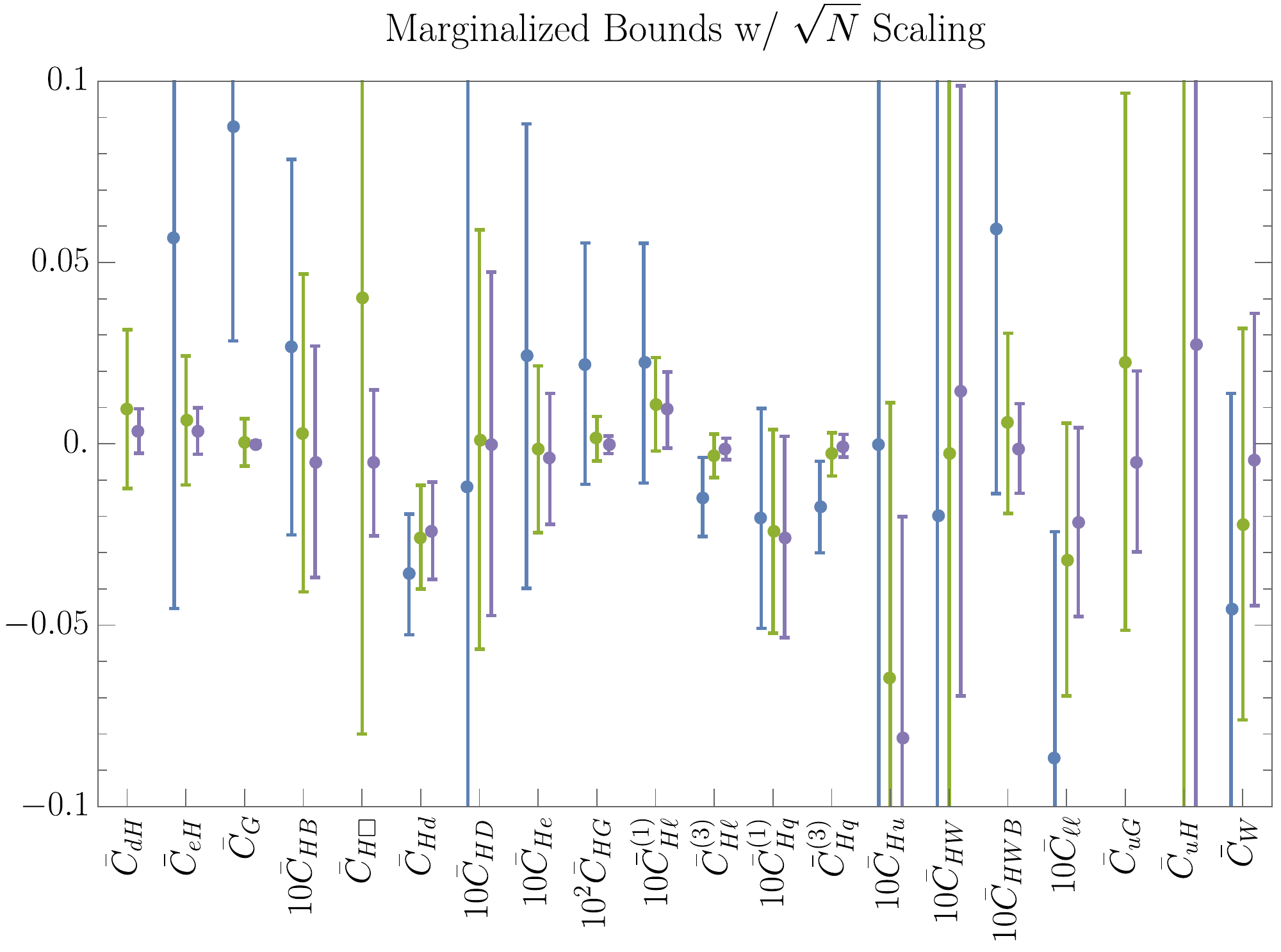}
  \includegraphics[width=0.45\textwidth]{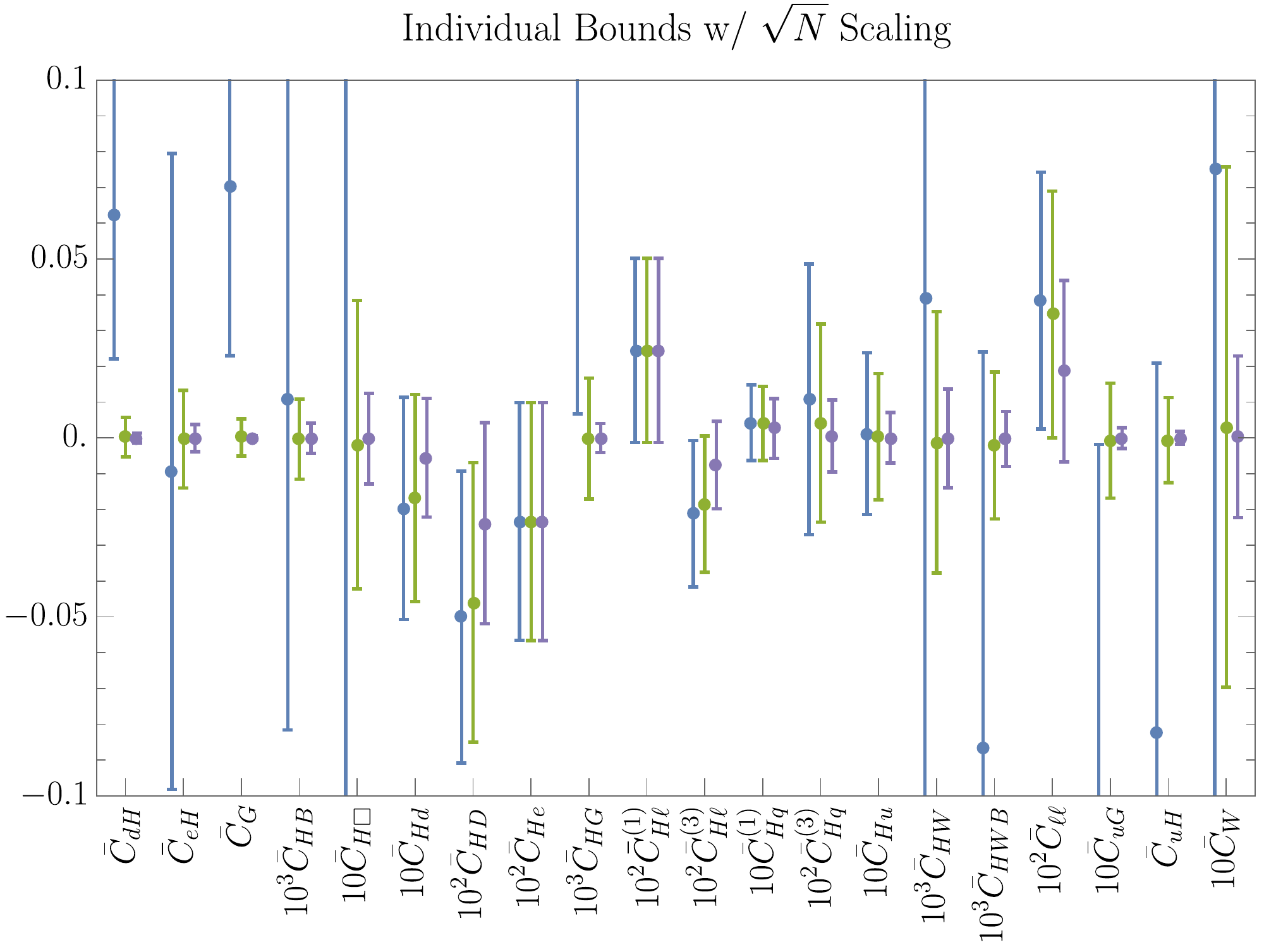}
 \caption{\it The best-fit values and 95\% CL ranges for dimension-6 operator coefficients using current data (blue), 
 and projections with the scaling (\ref{scaling}) for HL-LHC (green) and HE-LHC (purple) 
 measurements, including all operators simultaneously (left panel) and switching each operator on individually (right panel)~\cite{EMSY+}. }
   \label{fig:compHLHE}
\end{figure}

As a first attempt to project how these constraints might evolve in the future at the HL- and HE-LHC~\cite{EMSY+}, we have assumed the following
simple scaling laws for the experimental, systematic and theoretical uncertainties for each operator coefficient:
\begin{equation}
\label{scaling}
\frac{\delta\mathcal{O}_{\text{HL}}}{\delta\mathcal{O}_{\text{today}}} = \sqrt{\frac{L_{\text{today}, i}}{L_{\text{HL}}}} \; , \qquad
\frac{\delta\mathcal{O}_{\text{HE}}}{\delta\mathcal{O}_{\text{today}}} = \sqrt{\frac{\sigma_{13}}{\sigma_{27}} \frac{L_{\text{today}}}{L_{\text{HE}}}} \; .
\end{equation}
On the one hand this is optimistic, since there is no analysis supporting such scaling laws for the systematic and theoretical uncertainties,
but on the other hand this is pessimistic, since the higher luminosities and energy will make possible more refined measurements
of the kinematic distributions. We assume $L_{\text{today}} \simeq 36~\text{fb}^{-1}$ for most current LHC measurements,
and the benchmark luminosities $L_{\text{HL}} = 3~\text{ab}^{-1}$. and $L_{\text{HE}} = 15~\text{ab}^{-1}$ for all the HL- and HE-LHC
measurements. The results for the operator coefficients are shown in Fig.~\ref{fig:compHLHE},
where the current LHC constraints are shown in blue, 
and the prospective HL- and HE-LHC constraints are in green and purple, respectively.

\begin{figure}[h!]
\begin{center} 
\includegraphics[scale=0.4]{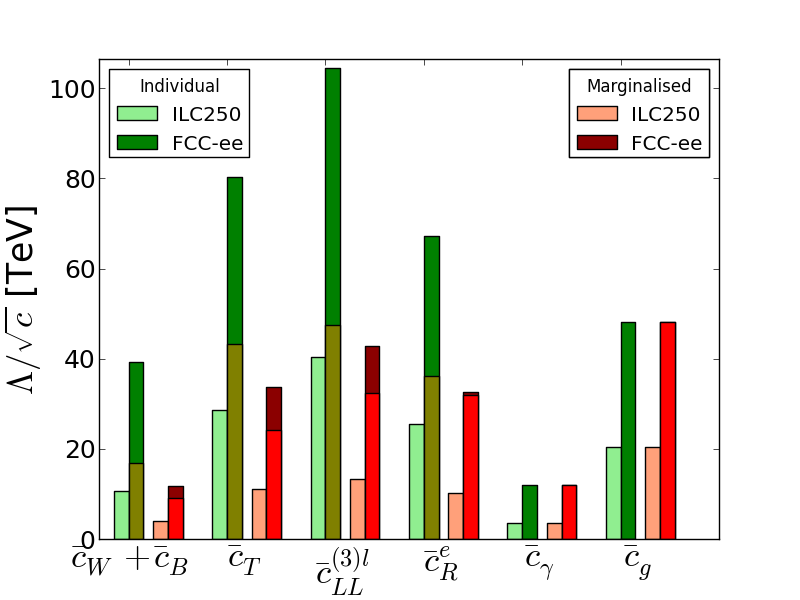}
\includegraphics[scale=0.4]{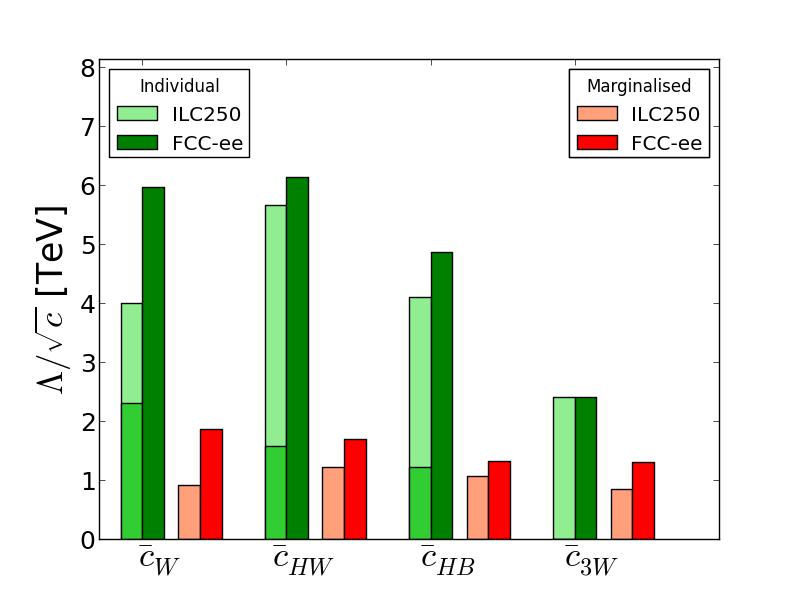}	
\caption{\it Summary of the reaches for the dimension-6 operator coefficients when switched on individually (green) 
and when marginalised (red), from projected precision measurements at the ILC250 (lighter shades) and 
FCC-ee (darker shades)~\cite{EY}. The left plot shows constraints from Higgs and electroweak data, with
the different shades representing the effects of theoretical uncertainties at FCC-ee. 
The right panel shows constraints from Higgs physics and triple-gauge couplings, 
and the different shades of light green show the effect of including the latter in the ILC analysis. }
\label{fig:e+e-}
\end{center}
\end{figure}

Results for fits to projected data 
from the ILC at energies $\le 250$~GeV and from FCC-ee are shown in Fig.~\ref{fig:e+e-}~\cite{EY}. In the left panel
we show constraints on the coefficients of operators most strongly constrained by Higgs production and electroweak precision data
in individual fits (green) and marginalized fits (red). In each case, the lighter (darker) shadings are for fits
with/without theoretical uncertainties in the electroweak measurements. In the right panel we show constraints on
operators contributing to Higgs and diboson production, again showing the results of individual fits in green and 
marginalized fits in red. In this case, we show in different shades the effects of including ILC data on triple-gauge
couplings. We see that the FCC-ee constraints are significantly stronger than those from the ILC, in particular
because of the greater accuracy in the electroweak precision measurements. We also note the importance of
controlling the theoretical uncertainties in the interpretation of the FCC-ee data.

\begin{table}[htb!]
\begin{center}
\begin{tabular}{|c|c|c|c|c|c|}
\hline
 & $\delta \Gamma_Z\;[\rm MeV]$ &  $\delta R_l \; [10^{-4}]$ & $\delta R_b\; [10^{-5}]$ & 
 $\delta \sin^{2,l}_{eff} \theta \; [10^{-6}]$
 \\ \hline
 \multicolumn{5}{|c|}{Present EWPO errors} 
\\
\hline
EXP1 & 2.3 & $250$ & $66$ & $160$ 
\\
 TH1   & 0.4  &  60   & $10$ & $45$
  \\  
 \hline 
\multicolumn{5}{|c|}{FCC-ee EWPO error estimates}  
\\
\hline
EXP2   & 0.1 & 10 &  $2\div 6$  & $6$
\\
\hline
TH2   & 0.15 & 15 & 5 & 15 \\
\hline
TH3   & $<0.07$ & $<7$ & $<3$ & $<7$ \\
\hline
\end{tabular}
\end{center}
\vspace{-2ex}
\caption{\it Comparison for selected precision observables of present experimental measurements (EXP1),
current theory errors (TH1), FCC-ee precision goals (EXP2), estimates of the theory errors assuming that 
electroweak 3-loop corrections are known (TH2), and assuming that also the 
dominant 4-loop corrections are available (TH3). Adapted from~\cite{FCC-ee-precision}.
\label{tab:errors}}
\end{table}

\begin{table}[htb!]
\begin{center}
\begin{tabular}[c]{|c|c|c|c|}
\hline
\multicolumn{4}{|c|}{{\large{ $Z \rightarrow b \bar{b}$}}}
\\
\hline
& 1 loop  & \hspace{0.9cm}2 loops\hspace{0.9cm} &  \hspace{1.2cm}3 loops \hspace{1.2cm} \\
Number of topologies
& 1 & 5   & 51\\
\hline
{Number of diagrams}& 15 & 1074 & 120472\\
\hline
{Fermionic loops} &0 & ${150}$ & $17580$\\
\hline
{Bosonic loops} &15 & ${924}$ & ${102892}$\\
\hline
{Planar / Non-planar} &15 / 0 & ${981 / 133}$ & ${84059 / 36413}$\\
\hline
{QCD / EW} &1 / 14&{98 / 1016}& ${10386 / 110086}$\\
\cline{2-4}
\hline
\end{tabular}
\end{center}
\caption{\it Numbers of topologies and diagrams  for $Z\to b \bar{b}$
decays in the Feynman gauge, using  topological
symmetries of diagrams, and after removing tadpoles and wave-function diagrams.
Adapted from~\cite{FCC-ee-precision}.} 
\label{tab:diagrams}
\end{table}

Considerable effort will be needed to reduce the SM theory uncertainties to the level where full
benefit can be extracted from the electroweak precision measurements at FCC-ee, in particular.
The proceedings of a workshop dedicated to assessing 
the magnitude of this task are available in~\cite{FCC-ee-precision}.
Table~\ref{tab:errors} lists, for selected electroweak precision observables, the current
experimental errors (EXP1), the current theoretical errors (TH1), the prospective measurement
errors at FCC-ee~\cite{FCCee} (EXP2), the estimated theoretical uncertainties if complete 3-loop calculations are
available (TH2), and the estimated theoretical uncertainties if the dominant 4-loop diagrams
are also calculated (TH3). Table~\ref{tab:diagrams} lists, for $Z \to b \bar b$ decay, the numbers of
distinct topologies of diagrams to be calculated, the numbers of diagrams and various categories
at the 1-, 2- and 3-loop order. The general opinion of the authors of~\cite{FCC-ee-precision}
were that, while daunting, these calculations should be feasible on the necessary time-scale.

\begin{figure}[htb!]
\begin{center}
\includegraphics[scale=0.4]{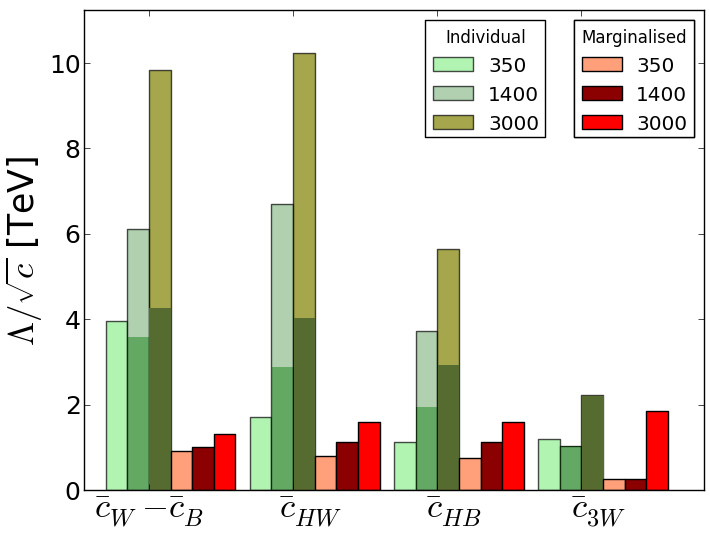}
\includegraphics[scale=0.41]{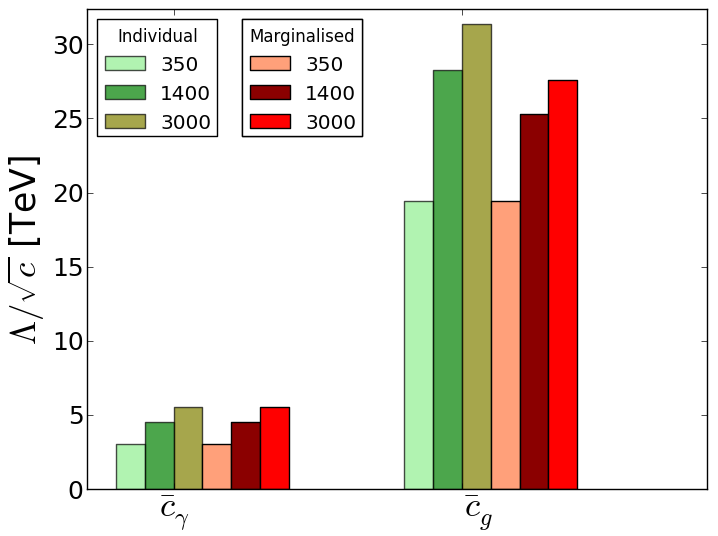}
\caption{\it  The estimated sensitivities of CLIC measurements at 350~GeV, 1.4~TeV and 3.0~TeV to the scales
of various dimension-6 operator coefficients, showing individual (marginalised) fit results as green (red) bars~\cite{ERSY-CLIC}. 
The lighter (darker) green bars in the left panel
include (omit) the prospective $HZ$ Higgsstrahlung constraint.}
\label{fig:CLIC}
\end{center}
\end{figure}

Fig.~\ref{fig:CLIC} shows the prospective sensitivities of CLIC measurements
at 350~GeV, 1.4~TeV and 3.0~TeV to the coefficients of various dimension-6 operators~\cite{ERSY-CLIC}.
As previously, the green bars are for fits in which each operator is switched on
individually, whereas the red bars are for fits including all operators. Also,
the lighter (darker) green bars in the left panel
include (omit) the prospective $HZ$ Higgsstrahlung constraint.
As could be expected, since the effects of higher-dimensional operators increase
with energy, the sensitivities are generally enhanced at higher centre-of-mass energies.

\section{Supersymmetry}

This (still) my favourite scenario for physics beyond the SM, despite the disappointment
that it has not (yet) been discovered at the LHC. Indeed, I would argue that Run 1 of the
LHC has, in addition to the traditional motivations of improving the naturalness of the
mass hierarchy, its role in string theory and its provision of a dark matter candidate,
provided three new reasons for liking supersymmetry. One is that it would stabilize the
electroweak vacuum~\cite{ER}, another is that it predicted correctly the mass of the Higgs boson~\cite{mh},
and the third is that predicted correctly that the Higgs couplings should resemble those
in the SM~\cite{EHOW}.

\begin{figure*}[htb!]
\centering
\includegraphics[width=0.49\textwidth]{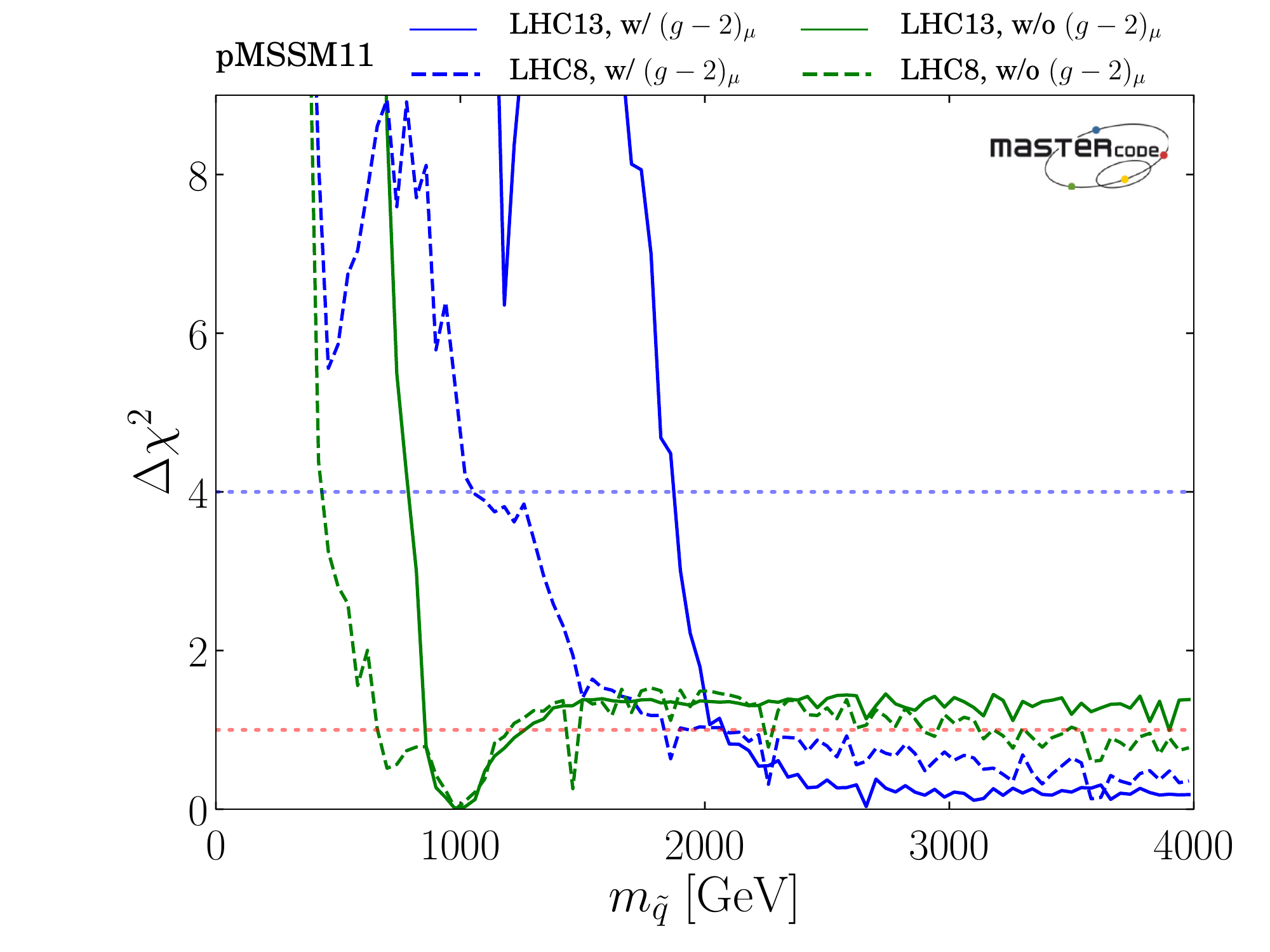}
\includegraphics[width=0.49\textwidth]{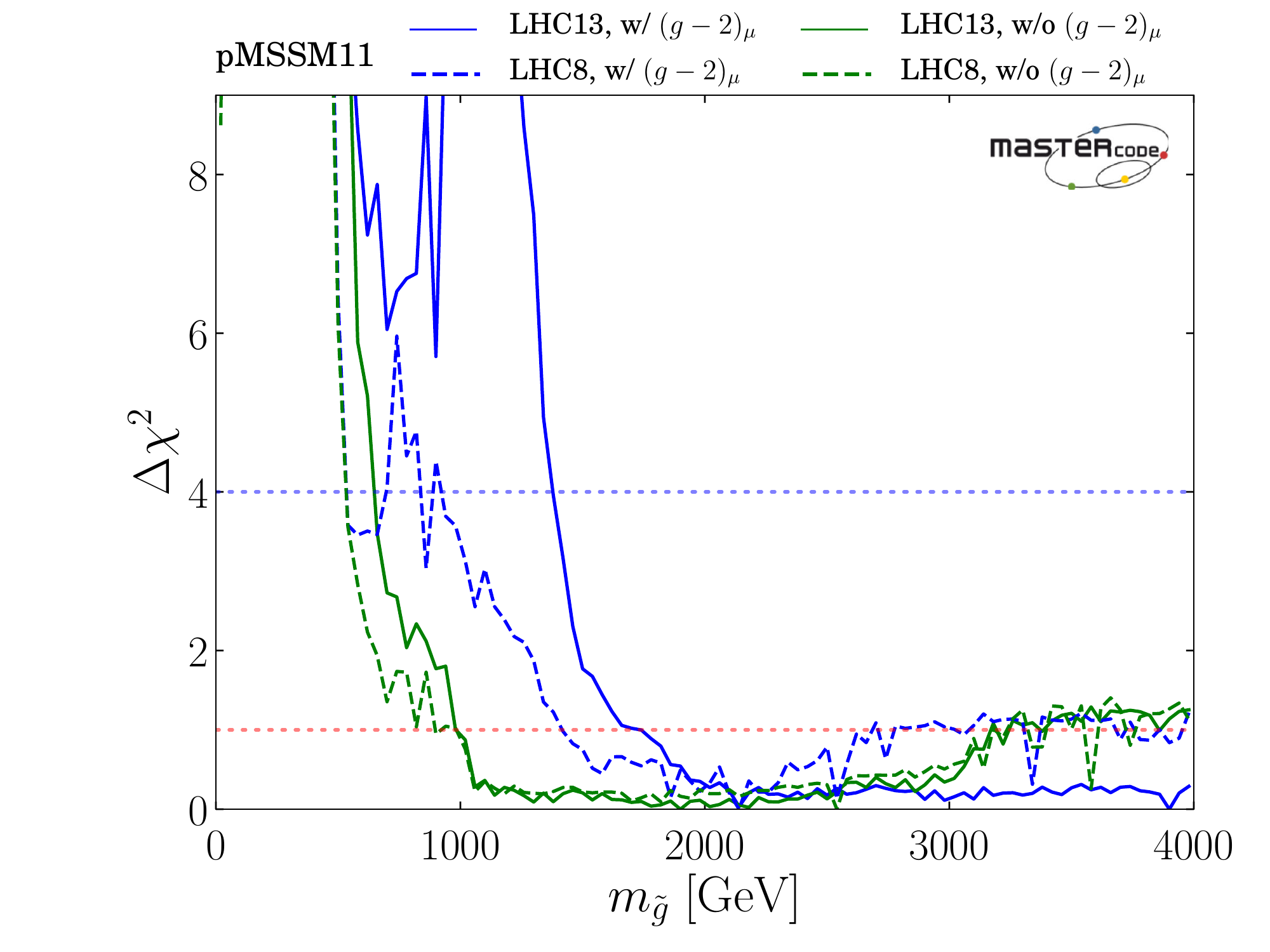} \\
\caption{\it One-dimensional $\chi^2$ functions for the 
squark mass (left panel) and the gluino mass (right panel) in the pMSSM11 with (blue) and
without the $g_\mu - 2$ constraint (green) and with (solid) and 
without (dashed) the 13-TeV LHC constraints~\cite{MCpMSSM11}.}
\label{fig:pMSSM11}
\end{figure*}

So where is supersymmetry to be found? Fig.~\ref{fig:pMSSM11} shows some results from recent global fits to 11
phenomenological parameters of the minimal supersymmetric extension of the SM (pMSSM11)
including the available experimental constraints including those from the first phase of
Run~2 of the LHC at 13 TeV~\cite{MCpMSSM11}. We see that both squarks and gluinos
could be as light as $\sim 1$~TeV if $g_\mu - 2$ is dropped from the fit, whereas somewhat
heavier masses are preferred if $g_\mu - 2$ is included in the fit. Either way, strongly-interacting
sparticles could well lie within reach of future LHC runs.

\begin{figure*}[htb!]
\centering
\includegraphics[width=0.9\textwidth]{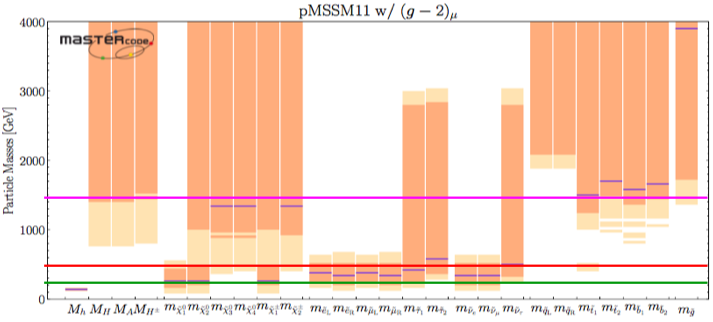}
\caption{\it Sparticle spectrum for the pMSSM11 with the $g_{\mu} -2$ constraint applied~\cite{MCpMSSM11}.
The masses at the best-fit points are indicated by horizontal blue lines,
and the 68 and 95\% CL ranges by vertical orange bands.
The kinematic reaches for pair-production of various sparticle species at $e^+ e^-$
colliders with centre-of-mass energies of 500 GeV, 1 TeV and 3 TeV 
are shown as horizontal green, red and mauve lines, respectively.}
\label{fig:pMSSM11masses}
\end{figure*}

A more complete picture of the possible sparticle mass spectrum in the pMSSM11 in the case 
where $g_\mu - 2$ is dropped from the fit is shown in Fig.~\ref{fig:pMSSM11masses}~\cite{MCpMSSM11}. The
best-fit mass values are shown as horizontal blue bars, and the vertical orange bands show
the 68 and 95\% CL ranges for each of the sparticle masses. Also shown as horizontal lines
are the kinematic reaches for pair-production of various sparticle species at $e^+ e^-$
colliders with centre-of-mass energies of 500 GeV (green), 1 TeV (red) and 3 TeV (mauve).
The two former correspond to possible energy upgrades of the ILC (which is currently proposed
to reach 250 GeV), and the third line corresponds to CLIC. We see that a 500-GeV collider
would have little chance of producing sparticles in the pMSSM11, whereas a 1-TeV collider
would have a better chance, and CLIC could have very interesting prospects for detecting and
measuring sparticles - we will know better at the end of LHC Run 2.
 
\begin{figure*}[htbp!]
\centering
\includegraphics[width=0.49\textwidth]{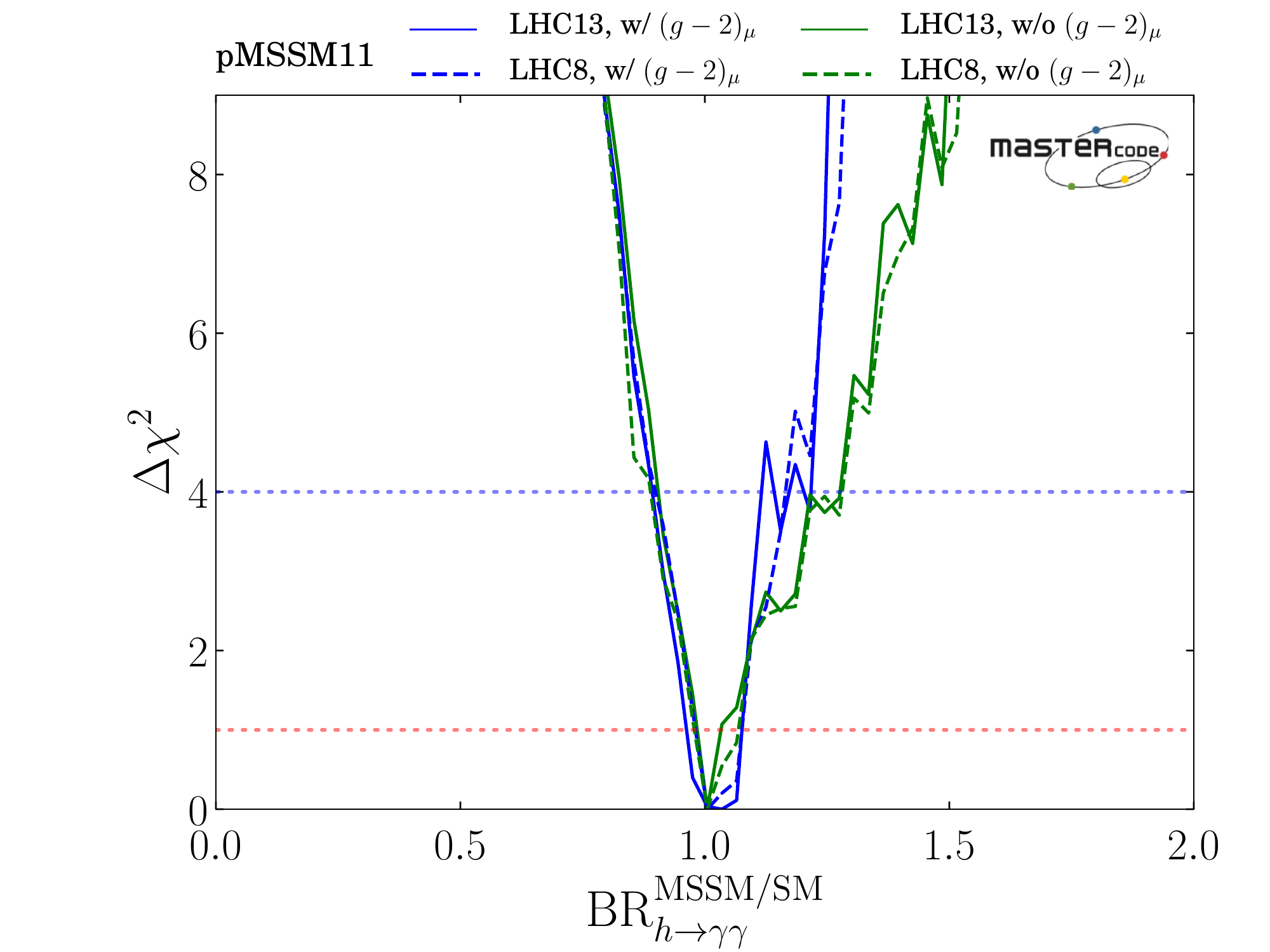}
\includegraphics[width=0.49\textwidth]{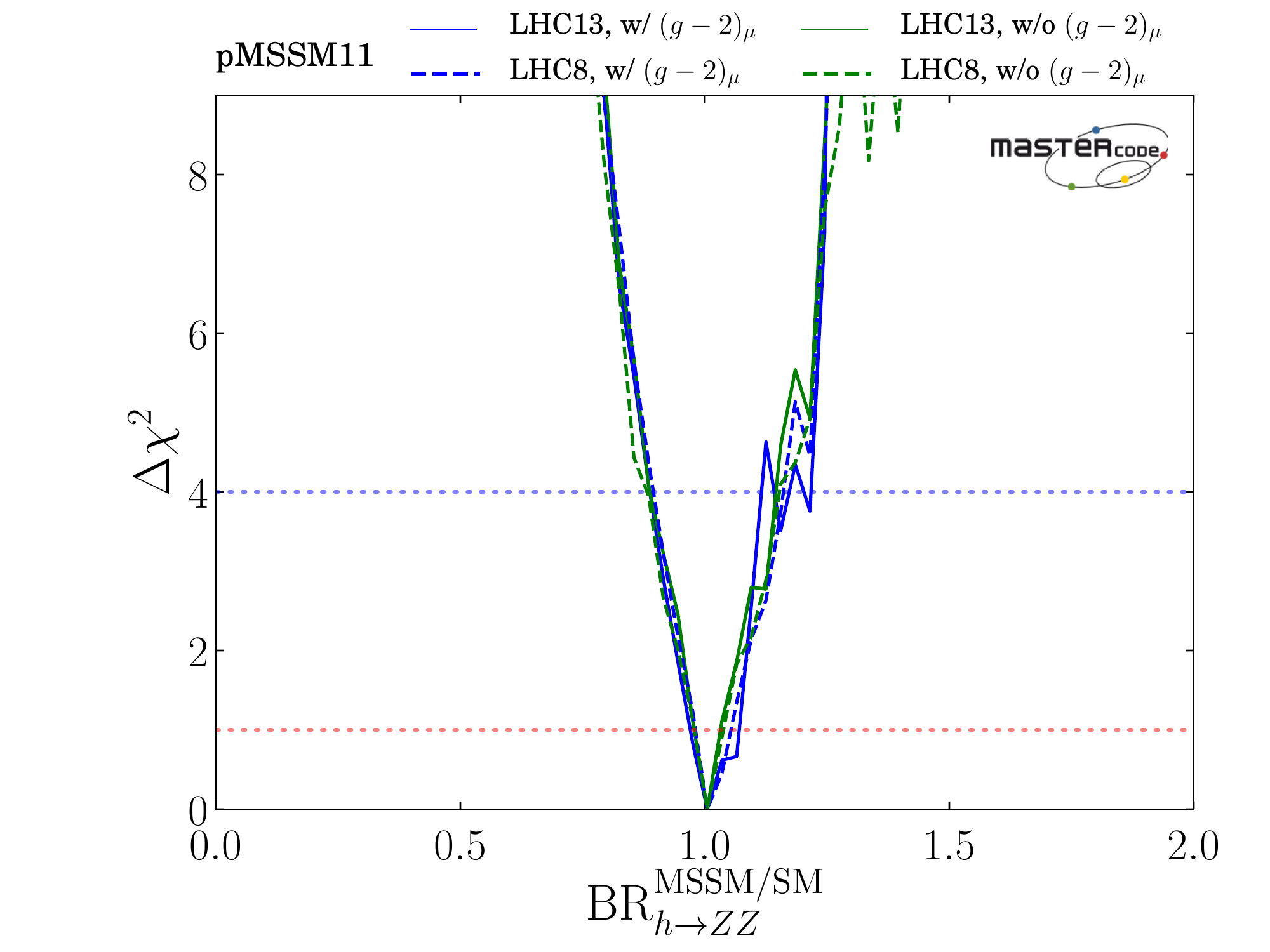}
\caption{\it One-dimensional $\chi^2$ functions for the branching ratios
for $h \to \gamma \gamma$ (left panel) and $h \to Z Z^*$ (right panel)  in the pMSSM11 fits with (blue) and
without the $g_\mu - 2$ constraint (green) and with (solid) and 
without (dashed) the 13-TeV LHC constraints~\cite{MCpMSSM11}.}
\label{fig:1hBR}
\end{figure*}

The global fits of the pMSSM11 with and without $g_{\mu} -2$ and 13-TeV LHC data are all
highly compatible with the measured mass of the Higgs boson. Also, as seen in Fig.~\ref{fig:1hBR}~\cite{MCpMSSM11},
they prefer values of the branching ratios for $h \to \gamma \gamma$ and $h \to Z Z^*$ that are
similar to those in the SM. However, they also allow for substantial deviations $\lesssim 20$\%
at the 95\% CL. Therefore, low-energy $e^+ e^-$ colliders such as ILC250, FCC-ee or
CEPC may still have good prospects for discovering indirect signatures of supersymmetry,
even though they may have little chance of producing sparticles directly.

\begin{figure*}[htbp!]
\centering
\includegraphics[width=0.44\textwidth]{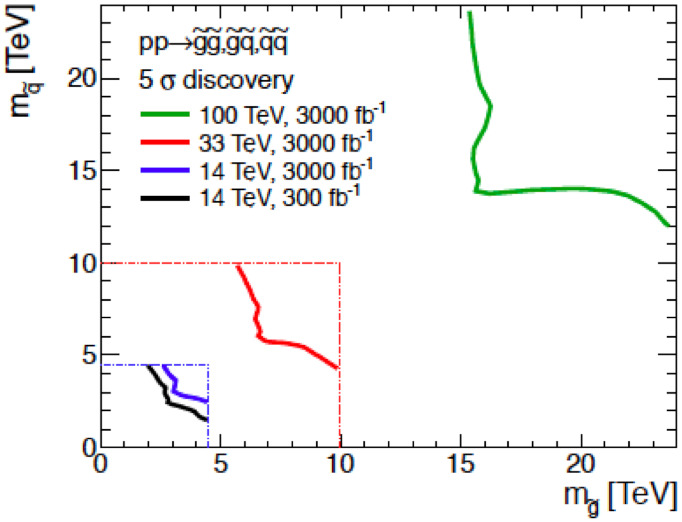}
\includegraphics[width=0.54\textwidth]{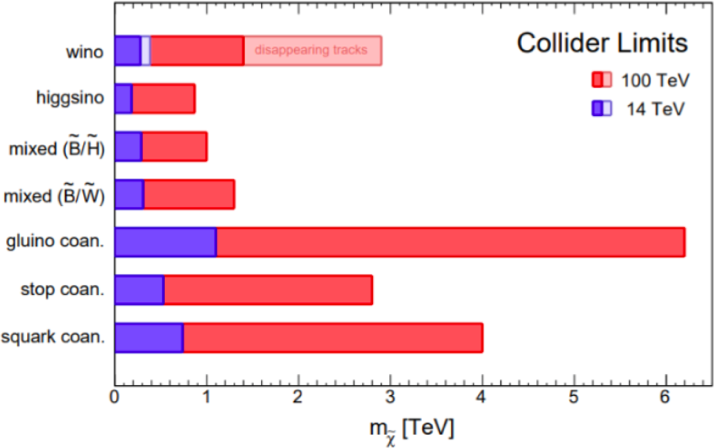}
\caption{\it The reaches at a 100-TeV $pp$ collider for conventional missing-energy searches for squarks and gluinos
(left panel) and different neutralino candidates (right panel)~\cite{FCC-hh}.}
\label{fig:susy100}
\end{figure*}

What are the prospects for sparticle discovery at a 100-TeV $pp$ collider? The left panel of Fig.~\ref{fig:susy100}
shows that the reach for conventional missing-energy searches for squarks and gluinos extends
well above 10 TeV~\cite{FCC-hh}, as in the vision outlined earlier. Moreover, the reach for various different neutralino
candidates extends into the multi-TeV range, as seen in the right panel of Fig.~\ref{fig:susy100}.

\begin{figure}[hbtp!]
\centering
\includegraphics[height=6cm]{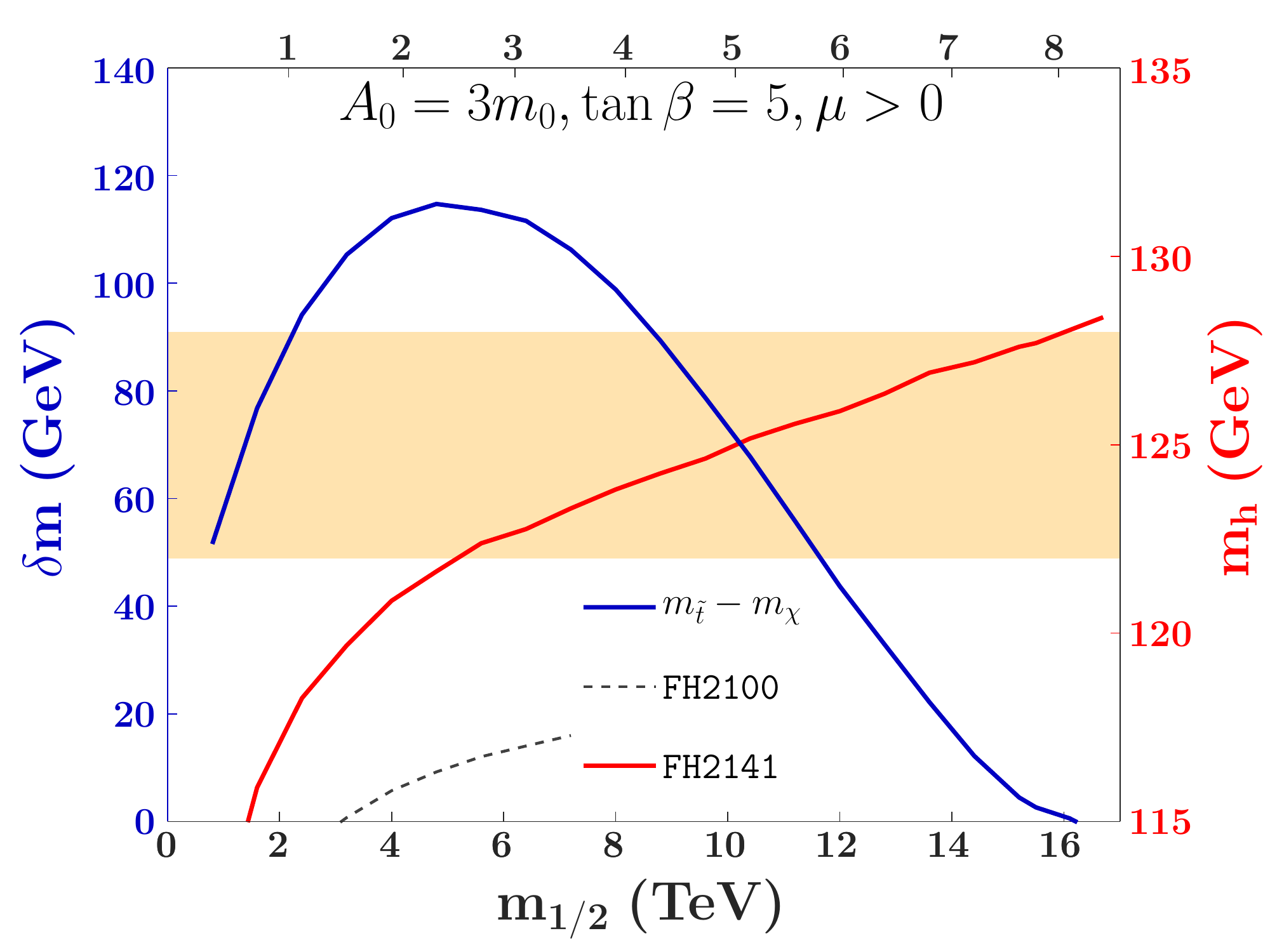}
\caption{\it
The profile of the stop coannihilation strip in the constrained MSSM
for $\tan \beta = 5, A_0 = 3 m_0$ and $\mu > 0$~\cite{EHOW++}.
The lower horizontal axis shows $m_{1/2}$ and the upper horizontal axis shows the corresponding values of 
the lightest neutralino mass.
The blue curve shows the stop-neutralino mass difference, to be read from the left vertical axis.
The red line shows the value of $m_h$ calculated using the latest version {\tt 2.14.1} of the {\tt FeynHiggs} code~\cite{FH},
to be read from the right vertical axis. Calculations within the orange band may be regarded as acceptable,
given the uncertainties in the calculations.}
\label{fig:stopstrip}
\end{figure}

Does this mean that there is a no-lose theorem for discovering supersymmetry at a 100-TeV $pp$
collider? Unfortunately not. Arguments about the possible sparticle mass scale based on the naturalness
of the mass hierarchy, grand unification or the stability of the electroweak vacuum do not set very
stringent upper limits on the scale of supersymmetry breaking. The tightest constraints known to me come from
the density of dark matter, assuming that the lightest sparticle is a stable neutralino. An order-of-magnitude
argument suggests that the neutralino should weigh ${\cal O}$(TeV) in order to avoid providing too much
dark matter. However, as seen in the right panel of Fig.~\ref{fig:susy100}, there are circumstances in which
the neutralino could be significantly heavier, specifically when it is almost degenerate with some other
sparticle(s), and coannihilations suppress the relic neutralino density. Scenarios that have been studied
in some detail include gluino and stop coannihilation and Fig.~\ref{fig:stopstrip} shows one example of the
profile of a stop coannihilation strip in the constrained MSSM~\cite{EHOW++}. The stop-neutralino mass difference $\delta m$
vanishes when the the supersymmetry-breaking parameter $m_{1/2} \to 16$~TeV, corresponding to
neutralino and stop masses $\simeq 8$~TeV. As shown by the red line, the value of $m_h$ calculated using
the latest version of the {\tt FeynHiggs} code is compatible with the experimental value, within the
calculational uncertainty represented by the horizontal orange band. This example is certainly quite
finely tuned, but it does indicate that new ideas~\cite{EZ} may be needed to be sure of discovering sparticles even at a 100-TeV
$pp$ collider.

\section{Conclusions}

There are many reasons to expect physics beyond the SM, including the apparent instability of the
electroweak vacuum calculated within the SM, the nature of dark matter, the origin of matter,
the naturalness of the hierarchy of mass scales, neutrino masses and mixing, the mechanism of cosmological
inflation, and constructing a quantum theory of gravity. The strongest arguments for new physics at the TeV
scale that could be accessible to future high-energy colliders are provided by dark matter and the naturalness problem.

The HL-LHC is on its way, and will take Higgs studies
and the search for new particles to the next level, with almost two orders of magnitude more LHC data than have
been analyzed so far. There are also many ideas for possible new high-energy accelerators. The ILC may be
next, while CLIC is a project for a linear $e^+ e^-$ collider at higher energy, which has advantages for both
direct and indirect searches for new physics. Future circular $e^+ e^-$ colliders, on the other hand, are limited in the
centre-of-mass energy they can reach, but offer larger luminosities at low energies end hence a new generation
of high-precision electroweak tests. Moreover, a large circular tunnel also offers the prospect of $pp$ collisions at
energies as high as 100 TeV, as well as the possibility of $ep$ collisions, representing, in my view, the most versatile way forward.

\section*{Acknowledgment}

This work was supported in part by the United Kingdom STFC Grant ST/P000258/1, and in part by the Estonian Research Council via a Mobilitas Pluss grant.

\end{document}